# LUT-Assisted Clock Data Recovery and Equalization for Burst-Mode 50-100 Gbit/s Bandwidth-Limited Flexible PON

**Yanlu Huang[1], Liyan Wu[1], Shangya Han[1], Kai Jin[1], Kun Xu[1], Yanni Ou[1]***

*[1] State Key Laboratory of Information Photonics and Optical Communications, Beijing University of Posts and Telecommunications, Beijing 100876, China, *yanni.ou@bupt.edu.cn*

**Abstract:** We demonstrated LUT-assisted CDR and equalization for burst-mode 50-100 Gbit/s bandwidth-limited PON, achieving signal recovery under large 100 ppm frequency offsets and 0.5 UI phase mismatch using reduced 50ns preambles, with 0.3dB sensitivity penalty only. 

## 1. Introduction

ITU-T has recently published the recommendation for passive optical networks (PON) using the fixed NRZ modulation format [1]. To further extend the network capacity while also yielding the legacy ITU-T PON power budget requirements, flexible PON technologies (FLCS-PON) were proposed to provide ONUs with flexible modulation formats, coding and digital signal processing (DSP) options to better accommodate various channel conditions [2-4]. However, these works focused mainly on downstream (DS), while upstream (US) transmission has not yet been well considered. In contrast to the TDM-PON DS, the US transmission has more challenges in the burst-mode receiver design in the OLT. These challenges are mainly from the more stringent requirements in the faster adaptions and re-configurations of clock data recovery (CDR) and compensation for bursts consisting of flexible modulation formats. Considering the bandwidth-limitation nature of the flexible PON, the detailed challenges of the US signal reception include the following aspects.

First, inter-symbol interference (ISI) caused by bandwidth limitation and dispersion will result in severe signal distortions, requiring the signal recovery scheme to be trained with larger and longer training data. However, the size of the training data is required to be small since the bursts could be as short as few tens of bytes [5-6]. In addition, each burst in the flexible PON can carry different modulation formats, requiring fast re-configuration of the equalizer. A pre-calculated method has been proposed to achieve rapid convergence [7-9], while its application in the flexible PON US has not yet been verified. Apart from the signal recovery, quick CDR designing in burst-mode receiver is another challenge, as the bursts arriving at OLT have different powers and phases. Such a variety greatly increases the difficulties in achieving the CDR fast locking. The clock phase caching method was proposed to address the fast lock difficulty [10]. However, the impact of clock frequency offset (CFO) on CDR performance, which is crucial for practical implementation of this method in PON, has not yet been well considered.

In this paper, for the first time, the 50 Gbit/s and 100 Gbit/s bandwidth-limited signal recovery performance in a flexible TDM-PON US scenario based on the 25G-class optoelectronic devices is experimentally demonstrated and evaluated, by employing both the look-up table (LUT)-assisted CDR and LUT-assisted equalization schemes in combination. Results indicate the resilience and the robustness of the signal recovery performance using the proposed LUT-assisted CDR and LUT-assisted equalization schemes, under the clock frequency offset (CFO) conditions as large as 100 ppm. After using the LUT-assisted scheme, the convergence time to further locate the optimized CDR sampling point is largely reduced to as short as 50 ns, which is equivalent to the use of an additional 2500-symbol preamble for convergence. Moreover, compared to the system with no CFO and phase mismatch (PM), the receiver sensitivity shows less than 0.3 dB penalty even under 100 ppm CFO and 0.5 UI (unit interval) PM, showing the recovery stability in dealing with a large range of clock frequency offsets and phase mismatches.

## 2. LUT-assisted CDR and equalization scheme

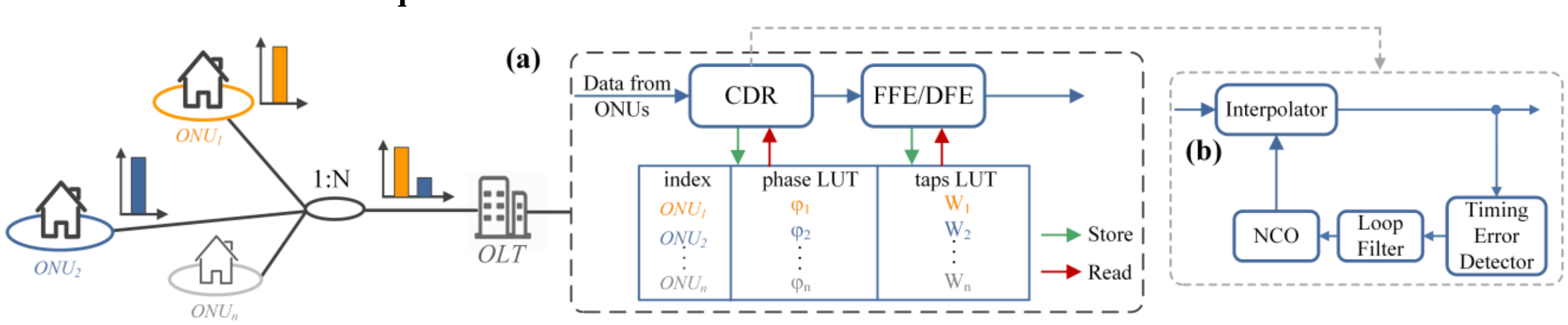

Fig. 1. The principle of (a) the proposed LUT-assisted CDR and LUT-assisted equalization scheme, and (b) the commonly adopted CDR module.

Fig. 1(a) shows the LUT structure of the proposed CDR and equalization schemes, while Fig. 1(b) represents the commonly adopted CDR method used in our experiment. Considering the PON US scenario, signals with different optical powers were considered to emulate the signals sent from different ONUs, as the optical power is almost linearly related to the distance between ONUs and OLT. As illustrated in Fig. 1(a), each ONU sends a signal to the OLT at the beginning; the OLT receives the signal and feeds it into the CDR for sampling at the optimal point. After the convergence of CDR, the clock phase for the ONU is stored in the phase LUT, together with the ONU-ID. Then, the output of the CDR is fed into the feedforward and decision feedback equalizer (FFE/DFE) for distortion compensation. After the convergence of the equalizer, the tap coefficients are stored in the taps LUT. Subsequently, when the next burst comes from the ONU, the OLT recognizes the ONU-ID of the burst and reads the corresponding stored phase and coefficients. It uses them as the initial phase for the CDR and the initial tap coefficients for the equalizer, respectively, thereby recovering the signal in a shorter iteration. The combination of both the LUT-assisted CDR and LUT-assisted equalization can enhance the system in tolerating the clock phase deviations, enabling stable signal recovery as the LUT-assisted equalizer can recover the data that is not yet fully synchronized.

### 3. Experimental setup and results analysis

Fig. 2 shows the experimental setup and the DSP procedures at both the transmitter and the receiver sides. The system 3dB and 6dB frequency response bandwidth is depicted in the inset (a) as 9.2 GHz and 16.2 GHz, where inset (b) details the system parameters. To emulate the flexible PON burst-mode transmission, ONU1, set 5 km farther from OLT, sends 50 Gbit/s NRZ signals, while ONU2, closer to OLT, sends 100 Gbit/s PAM4 signals. Both ONUs use the same type of MZMs and EAs from different batches, with 1310 nm lasers for optical carriers generation. PRBS^15 data is generated with pre-emphasis for both ONUs, where ONU1 employs 9-tap FFE and 3-tap DFE (9FFE&3DFE), and ONU2 employs 13FFE&5DFE. The signals are then shaped by a root-raised-cosine filter and are resampled before being fed into the AWG, then amplified and loaded onto the MZMs, with output optical powers set to ~3 dBm. The signal from ONU1 is transmitted over 5 km of SMF and combined with ONU2's signal in an optical coupler. The combined signals are transmitted through 20 km SMF, emulating typical PON distances. At OLT, a VOA is applied to adjust the received optical power (ROP), and a PIN+TIA with an SOA preamplifier is employed to detect the optical signal. Then, the signal is captured by an oscilloscope and processed by the DSP, including resampling, LUT-assisted CDR and equalization, and BER calculation sequentially. For LUT-assisted equalization, 9FFE&3DFE is used for NRZ and 19FFE&5DFE is used for PAM4.

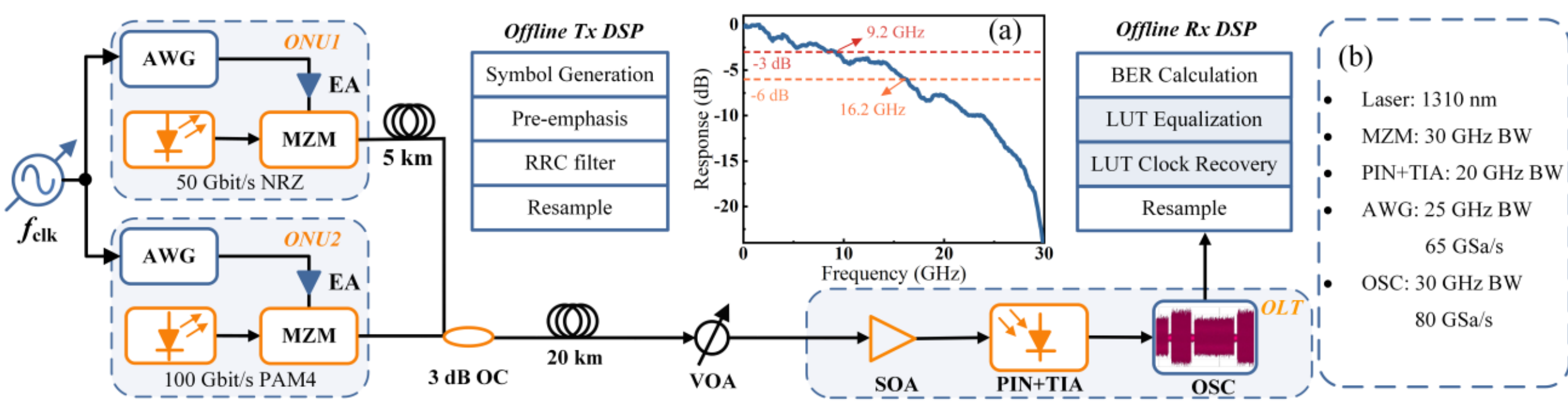

Fig. 2. Experimental setup for flexible PON burst-mode transmission and corresponding DSP procedures, with insets (a) system response, (b) component parameters. AWG: arbitrary waveform generator; MZM: Mach-Zehnder modulator; EA: electronic amplifier; OC: optical coupler; VOA: variable optical attenuator; SOA: semiconductor optical amplifier; TIA: transimpedance amplifier; OSC: oscilloscope.

To evaluate the performance of the LUT-assisted CDR and LUT-assisted equalization, the convergence time of the CDR and equalizer, as well as the BER performance under different CFOs and PMs are analyzed on the testbed. Fig3. (a)-(d) shows the CDR and equalization convergence with and without the LUT scheme. The normalized loop bandwidth, which impacts the CDR convergence time, is set to 3E-4 to ensure the synchronization precision while maintaining the locking speed. The condition that no frequency offset and no phase mismatch in the system is firstly evaluated. The phase mismatch refers to the discrepancies between the LUT phase values and the timing phase of the signal, caused by shifts in signals arriving at the OLT due to environmental variations. For the convergence of CDR, the CDR requires ~3700 symbols to converge without the LUT as show in Fig. 3(a), while it can precisely synchronized with the LUT as shown in Fig. 3(b). For the convergence of the equalizer, the equalizer requires ~3000 symbols for convergence without LUT, while the LUT-assisted equalizer needs no preambles to achieve this (Fig. 3(c) and (d)). Accordingly, the total number of ~6700 symbols is required for both CDR and equalizer convergence.

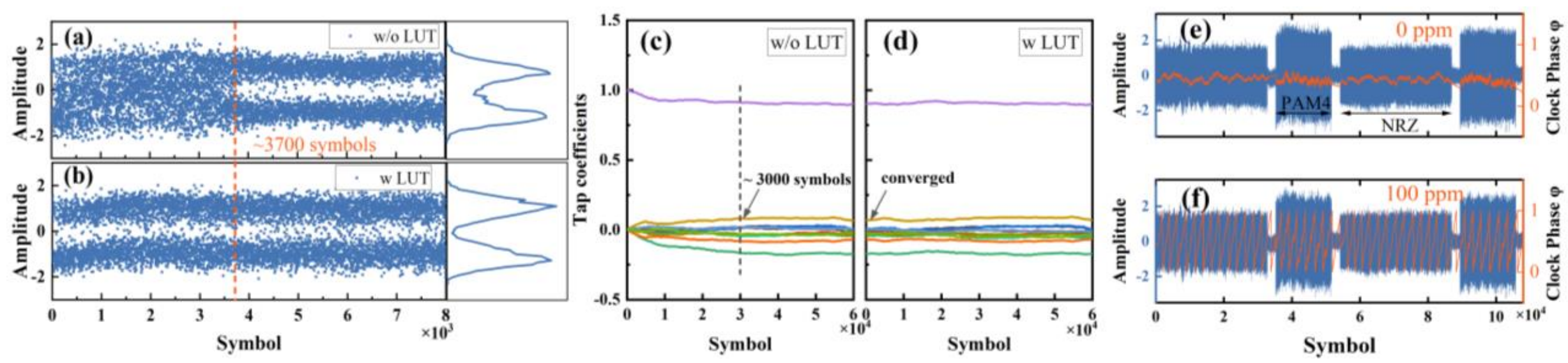

Fig. 3. (a) and (b) are the received NRZ signal after the CDR without and with LUT assistance, respectively. (c) and (d) are tap coefficient convergence curves of NRZ with and without LUT assistance. (e) and (f) are received bursts and estimated clock phases with LUT-assisted CDR at 0 ppm and 100 ppm CFO respectively.

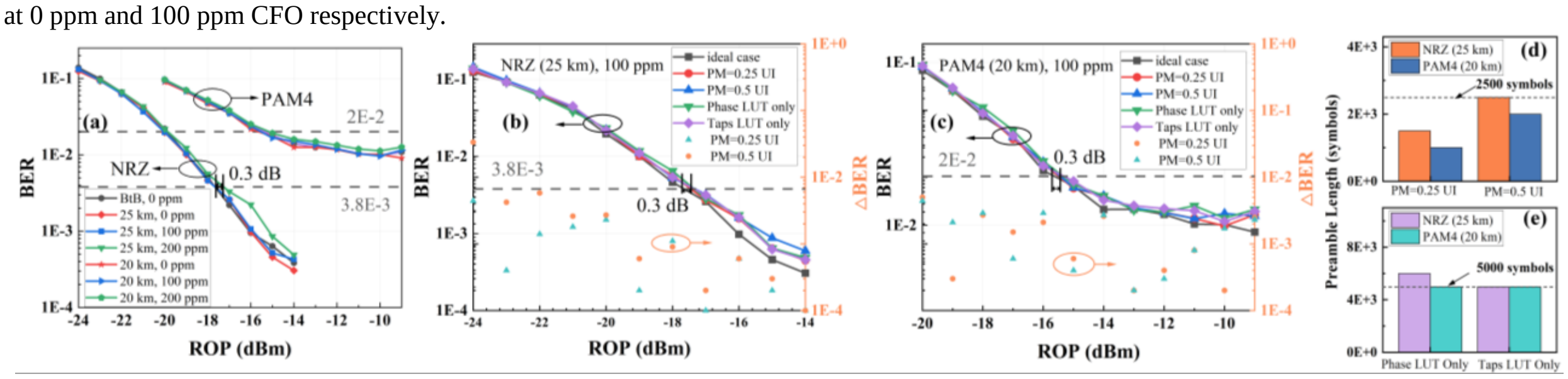

Fig. 4. BER performance after LUT-assisted CDR and LUT-assisted equalization. (a) BER at varying CFOs without phase mismatch. (b) and (c) are BER for NRZ (25 km) and PAM4 (20 km) at different phase mismatches with 100 ppm CFO, respectively. (d) Preamble length for CDR convergence at phase mismatch of 0.25 and 0.5 UI. (e) Preamble length for CDR convergence with only the phase LUT or only the taps LUT.

The condition that includes various CFOs and PMs in the system is then evaluated, in order to evaluate the resilience of the combination LUT scheme. For CFO testing, three scenarios are considered: the ideal 0 ppm, the challenging 100 ppm, and the stringent 200 ppm case. As shown in Figs. 3(e) and (f), the CDR locks onto the signal at the start of bursts and transitions the clock phase smoothly to accommodate timing variations in both 0 ppm and 100 ppm cases. The corresponding BER after applying the proposed combination scheme is presented in Fig. 4(a), which also includes the 200 ppm case. Notably, without using preambles for convergence, there is no performance deterioration for NRZ and PAM4 at 100 ppm. However, at 200 ppm, NRZ experiences a slight sensitivity penalty of 0.3 dB at the 3.8E-3 FEC threshold, due to timing phase deviations in the CDR. For PM testing, both extreme (0.5 UI) and moderate (0.25 UI) phase mismatch cases are evaluated, with the CFO set to 100 ppm. Fig. 4(b) and (c) show the BER results for NRZ (25 km) and PAM4 (20 km), and the sensitivity penalties for NRZ and PAM4 are 0.3 dB at 3.8E-3 and 2E-2, respectively. These were achieved with a 2500-symbol (50 ns) preamble for accommodating phase mismatch. This length is just 37.3% of that required by non-LUT-assisted schemes, and it is 50% of the minimum preamble length (5000 symbols) needed when using only the phase LUT or only the taps LUT under a 0.3 dB sensitivity penalty condition, as shown in Fig. 4(d) and (e).

## 4. Conclusion

We proposed and experimentally demonstrated a burst-mode 50/100 Gb/s bandwidth-limited flexible PON transmission system based on the combination of the LUT-assisted CDR and LUT-assisted equalization, enabling stable signal recovery at the receiver. The performance of the proposed scheme was experimentally verified for clock frequency offsets as large as 100 ppm. The results show that, compared to the ideal case without CFO and PM, our proposed scheme maintains a sensitivity penalty only within 0.3 dB under a relatively wide range of 100 ppm CFO and 0.5 UI PM. Comparing to the three scenarios: not using LUT, using only the phase LUT, and using only the tap LUT, the preamble length required for CDR convergence with the proposed LUT combine scheme can be significantly reduced by at least half.